\newcommand{\SEC}[1] {Section~\ref{#1}}  \newcommand{\FIG}[1] {Figure~\ref{#1}}

\newcommand{\SNR} {$S/N$}
\newcommand{\PT} {{$p_{\rm top}$}}

\newcommand{\PW} {{$w$}}
\newcommand{\IT} {{$\eta_{\rm T}$}}
\newcommand{\IF} {{$\eta_{\rm F}$}}
\newcommand{\ID} {{$\eta_{\rm DM}$}}
\newcommand{\PP} {PEACE}
\newcommand{\FIRSTL} {0.17\%}
\newcommand{\SECL} {0.34\%}
\newcommand{\THRL} {3.7\%}
\newcommand{\NPSR} {47}

\newcommand{\PAF}{PALFA}
\newcommand{\GBN}{GBNCC}
\newcommand{\GBD}{GBT Drift}

\RequirePackage{lineno}
\setpagewiselinenumbers
\modulolinenumbers[2]
\documentclass[useAMS,usenatbib]{mn2e}
\usepackage{graphicx, amsmath}

\title[ PEACE: Pulsar Evaluation Algorithm for Candidate Extraction ]
{PEACE: \emph{Pulsar Evaluation Algorithm for Candidate Extraction} ---
A software package for post-analysis processing of pulsar survey candidates}

\author[K.~J.~Lee et al.]{
K.~J.~Lee $^{1}$\thanks{Email: kjlee@mpifr-bonn.mpg.de},
K.~Stovall $^2,^3$,
F.~A.~Jenet ${^2}$,
\newauthor
J. Martinez${^2}$,
L.~P.~Dartez${^2}$,
A. Mata${^2}$,
G. Lunsford${^2}$,
\newauthor
S. Cohen${^2}$,
C.~.M.~Biwer${^{18}}$,
M. Rohr${^{18}}$,
J. Flanigan${^{18}}$,
A. Walker${^{18}}$,
S. Banaszak${^{18}}$,
\newauthor
B.~Allen$^{4,18}$,
E.~D.~Barr$^{1}$,
N.~D.~R.~Bhat$^{5,15}$,
S.~Bogdanov$^6$,
A.~Brazier$^7$,
F.~Camilo$^{6,19}$,
\newauthor
D. J.~Champion$^{1}$,
S.~Chatterjee$^7$,
J.~Cordes$^7$,
F.~Crawford$^8$,
J.~Deneva$^{19}$,
G.~Desvignes$^1$,
\newauthor
R.~D.~Ferdman$^{13,12}$
P.~Freire$^{1}$,
J. W. T.~Hessels$^{10, 11}$,
R.~Karuppusamy$^{1}$,
V.M.~Kaspi$^{12}$,
\newauthor
B.~Knispel$^4$,
M.~Kramer,$^{1,13}$
P.~Lazarus$^1$,
R.~Lynch$^{12}$,
A.~Lyne$^{13}$,
M.~McLaughlin$^{14}$,
\newauthor
S.~Ransom$^{16}$,
P.~Scholz$^{12}$,
X.~Siemens$^{18}$,
L.~Spitler$^{1}$,
I.~Stairs$^{17}$,
M.~Tan$^{17}$,
\newauthor
J.~van Leeuwen$^{10, 11}$,
W.~W.~Zhu$^{17}$,\\
$^1${Max-Planck-Institut f\"ur Radioastronomie, Auf dem H\"ugel 69,
	D-53121 Bonn, Germany} \\
	$^2${Center for Advanced Radio Astronomy, University of Texas at Brownsville, 
	Brownsville, TX 78520, USA }\\
	$^3${Department of Physics and Astronomy, University of Texas at San Antonio, 
	San Antonio, TX 78249, USA}\\
	$^4${Max-Planck-Institut f\"ur Gravitationsphysik,D-30176 Hanover, Germany }\\
	$^5${International Centre for Radio Astronomy Research, Curtin University, 
	Bentley, WA 6102, Australia}\\
	$^6${Columbia Astrophysics Laboratory, Columbia University, NY 10027, USA}\\
	$^7${Astronomy Department, Cornell University, Ithaca, NY 14853, USA}\\
	$^8${Department of Physics and Astronomy, Franklin and Marshall College, 
	P.O.  Box 3003, Lancaster, PA 17604-3003, USA}\\
	$^{10}${ASTRON, the Netherlands Institute for Radio Astronomy, Postbus 2, 7990 
	AA, Dwingeloo, The Netherlands}\\
	$^{11}${Astronomical Institute ``Anton Pannekoek,'' University of Amsterdam, 
	Science Park 904, 1098 XH Amsterdam, The Netherlands}\\
	$^{12}${Department of Physics, McGill University, Montreal, QC H3A 2T8, 
	Canada}\\
	$^{13}${University of Manchester, Jodrell Bank Observatory, Macclesfield, 
	Cheshire, SK11 9DL, UK}\\
	$^{14}${Department of Physics, West Virginia Univ., Morgantown, WV 26506,USA 
	}\\
$^{15}${Centre for Astrophysics \& Supercomputing, Swinburne University, 
Hawthorn, Victoria 3122, Australia}\\
	$^{16}${NRAO, Charlottesville, VA 22903, USA}\\
	$^{17}${Department of Physics and Astronomy, University of British Columbia, 
	6224 Agricultural Road, Vancouver, BC V6T 1Z1,Canada}\\
	$^{18}${Center for Gravitation, Cosmology and Astrophysics, University of
Wisconsin Milwaukee, Milwaukee WI, 53211, USA}\\
$^{19}${Arecibo Observatory, HC3 Box 53995,
Arecibo, PR 00612, USA}\\
}
\begin{document}

\date{\today}
\pagerange{\pageref{firstpage}--\pageref{lastpage}} \pubyear{2010}
\maketitle
\label{firstpage}

\begin{abstract}

Modern radio pulsar surveys produce a large volume of prospective
candidates, the majority of which are polluted by human-created radio
frequency interference or other forms of noise. Typically, large numbers of 
candidates need to be visually inspected in order to determine if they
are real pulsars. This process can be labor intensive. In this
paper, we introduce an algorithm called PEACE (\emph{Pulsar Evaluation
Algorithm for Candidate Extraction}) which improves the efficiency of
identifying pulsar signals. The algorithm ranks
the candidates based on a score function. Unlike popular machine-learning based 
algorithms, no prior training data sets are required. This algorithm
has been applied to data from several large-scale radio pulsar surveys. Using 
the human-based ranking results
generated by students in the Arecibo Remote Command Center programme, the
statistical performance of \PP\ was evaluated. It was found that \PP\
ranked 68\% of the student-identified pulsars within the top \FIRSTL\
of sorted candidates, 95\% within the top \SECL, and 100\% within the
top \THRL. This clearly demonstrates that PEACE significantly increases
the pulsar identification rate by a factor of about 50 to 1000. To
date, PEACE has been directly responsible for the discovery of \NPSR\
new pulsars, 5 of which are millisecond pulsars that may be useful for
pulsar timing based gravitational-wave detection projects. \end{abstract}


\section{Introduction}

Radio pulsars are unique celestial objects that are used as probes
to study a wide range of physics and astrophysics (see, for example,
\citealt{BH94, LK05, LG06}). Studies of pulsar emission have shed
light on the properties of the interstellar medium and the physics of
ultra-relativistic plasmas under high magnetic field conditions. The
statistical properties of the pulsar population give us important
information on the late stages of stellar evolution, the equation of
state of exotic material, and the formation and evolution of binary
and multiple-star systems. The stable rotation of radio pulsars allows
for unique tests of gravitation theories as well as the positive detection of 
ultra
low-frequency gravitational waves. In all cases, the greater the number of  
pulsars that are
discovered, the more physics and astrophysics we are able to study.

As of 2013, more than 2000 pulsars have been found (ATNF Pulsar Catalogue, 
\citealt{MHT05}).  Since pulsar
population models predict that the number of detectable pulsars in the
Galaxy should be about 10 times higher than this (assuming a luminosity 
threshold
of $0.1\, {\rm mJy\, kpc^2}$, \citealt{FK06, LFL06}), several major radio
observatories around the world are conducting large-scale blind searches
for more of these objects. Typically, pulsar surveys are performed by
pointing the telescope at a region of the sky for several minutes to
hours, then moving to another region, and repeating. Sophisticated analysis
software packages, like PRESTO or SIGPROC \citep{Ransom01, Lor01}, are
applied to the data to search for periodic signals while taking into
account the effects of dispersion by the interstellar medium as well as
Doppler shifts due to the binary orbital motion. These search algorithms
produce a possible series of ``candidates'' (i.e. files or plots containing the
identified periodic signals and their properties). One can find detailed 
information about pulsar searching techniques from the standard references 
\citep{LK05,LG06}. Visual inspection,
usually aided by graphical tools \citep{KEL09, FSK04}, is still required to
determine if a particular candidate is indeed a pulsar, as opposed to
radio-frequency interference (RFI). After inspection, the good candidates
are re-observed in order to confirm their astrophysical origin.

Surveys produce millions of candidates. As an example, the \emph{North High 
Time Resolution Universe pulsar survey} (HTRU North),  being conducted with the 
100-m Effelsberg radio telescope, is expected to produce 14 million pulsar 
candidates \citep{Barr11}.  The multibeam
survey currently ongoing at the Arecibo Radio Observatory, known as the
\emph{Pulsar Arecibo L-band Feed Array survey} (\PAF, \citealt{CFL06}), 
generates over
half a million pulsar candidates per year. The \emph{Green Bank Northern
Celestial Cap pulsar survey} (\GBN), currently being performed at the
Robert C. Byrd Green Bank Radio telescope, produces candidates at about
the same rate. The \emph{Green Bank Telescope 350MHz Drift scan pulsar
survey} (\GBD, \citealt{BLR13, LB13}) generated over 1.2 million candidates.  
Assuming that it takes about one second to
inspect a candidate by eye, one needs over 250 person-hours to
evaluate one million candidates. The number of candidates is
beyond the ability of a single person. There are two natural ways to solve the 
problem: 1) apply more person power, 2) use computer-based methods to reduce the 
number of candidates needing inspection. 

In an effort to gather the necessary person power and increase the rate at
which pulsars are identified, the \emph{Arecibo Remote Command Center
program} (ARCC), developed at the University of Texas at Brownsville
(UTB), trains students to search through a large number of candidates
from these pulsar surveys. In order to increase the rate of pulsar
discoveries, the ARCC students have developed a pulsar viewing software
package known as \emph{ARCC Explorer}, which contains a set of web
applications that allows multiple users to visually inspect and rank pulsar candidates. Further
details of the ARCC Explorer will be described elsewhere (Stovall et
al. in preparation), while in this paper we focus on one part of the
ARCC Explorer that analyzes the candidates generated by pulsar search
pipelines and ranks them according to how ``pulsar-like'' they are. We
call this tool the \emph{Pulsar Evaluation Algorithm for Candidate Extraction} 
(\PP).

There are currently two major techniques used to reduce the amount of the
candidates to inspect. The first type is to select suitable candidates
based on several selection rules. For example, graphical tools \citep{FSK04,
KEL09} have been developed to help the visual selection. The second method 
\citep{KEL09, EAT09, LEE09, EMK11, BB12} is to use computers
to automatically select and rank the candidates. \PP, which is a tool of the 
second type, calculates a score for each of the candidates, where
the score is a measurement of the degree to which a
candidate matches certain pulsar-like features. Based on the score from
PEACE, the ARCC Explorer prioritizes and distributes the candidates to the
students for evaluation. In this way, the pulsar-like candidates are given
to the students earlier than the rest of the candidates. At the time of writing 
this paper,
\PP, has helped to identify a total of \NPSR\ new pulsars in \PAF, \GBN, and
HTRU North, five of which are millisecond pulsars.  Students in the
ARCC program at UTB and University of Wisconsin-Milwaukee (UWM) identified
\PAF\ and \GBN\ candidates as pulsars, and they were later re-observed and 
confirmed. 

This paper describes the details of the algorithms used by \PP\ and the
techniques used to evaluate its efficiency in prioritizing candidates. The 
details of the pulsars found by \PP\ will be discussed elsewhere. The
algorithms are described in \SEC{sec:met}. In \SEC{sec:imp}, the current
implementation is discussed in detail, together with instructions
on how to obtain and install \PP. In this section, we also discuss
the efficiency of the current implementation, which is evaluated by
comparing its ranking of \GBN\ survey candidates against the human-based 
ranking generated
by ARCC students. We discuss these results and conclusions are presented in 
\SEC{sec:con}.

\section{Method to rank candidates}
\label{sec:met}

In this section, we explain the algorithm implemented in \PP\ to
rank candidates. \PP\ has two major parts. First, it analyzes the
candidate files from pulsar search packages (e.g. PRESTO and SIGPROC)
and, second, calculates several statistics such as the signal-to-noise ratio, 
the
pulse profile width, etc. We refer to these statistics as \emph{quality
factors}. From these quality factors \PP\ computes a score, which is
then used to rank the candidates. We define the quality factors and
describe how to calculate them in \SEC{sec:qfac}. The pulsar ranking
technique is then presented in \SEC{sec:like}.

\subsection{Quality Factors}
\label{sec:qfac}

Empirical experience has shown that one needs to inspect several features of a 
pulsar candidate in order to properly characterize it. In \PP\, we have 
implemented six quality factors, which are described below. For further details 
of the implementation, we refer readers to the documents in the code repository 
\footnote{The software can be downloaded from 
``http://sourceforge.net/projects/pulsareace/''.}.  These scores inevitably 
introduce selection effects in the searching process; we delay the related 
discussions to \SEC{sec:con}. 

\begin{enumerate}

	\item The signal-to-noise (\SNR) ratio of the folded pulse profile.

The \SNR\ is a measure of the significance of the signal present in the data.
There are various definitions of \SNR. Here we define the \SNR\ in relation to 
the pulse profile, where the \SNR\ is the ratio between peak and 
root-mean-square (RMS) values. \PP\ reads in the pulse profile data, determines 
the peak amplitude from the pulse profile. To avoid biasing the RMS estimation 
by outliers, we sort the profile data by intensity values and exclude the top 
10\%.  The \SNR\ is calculated as the ratio between the amplitude and the RMS 
value.  Although such definition of \SNR\ depends on the number of bins used to 
fold the profile, where fewer bins give higher \SNR, we did not find 
significant correlation between \SNR\ and pulsar period in the \GBN\ data set.  
There are
other definitions for the \SNR\, e.g. the \SNR\ in terms of the mean flux, the 
reduced $\chi^2$, the standard deviation of profile etc. Any of these quantities 
can be used to quantify the strength of pulsed signals, because they contain 
similar information. But one needs to find appropriate score functions or 
weights, as we will discuss below. 

\item The topocentric period of the source (\PT).

Pulsar search codes are designed to detect a periodic signal. As a result, each 
candidate has an associated signal period, the value of which can be indicative 
of the signal's origin. For example, RFI signals due to air traffic
control radar typically have periods of a few seconds and RFI induced by power 
systems have characteristic frequencies of 50/60 Hz depending on the
geographical location of the telescope. \PP\ reads the period directly
from the candidate file.

\item The width of the pulse profile (\PW).

The pulse width, \PW\, is defined as the width of the pulse
normalized by the candidate period. Therefore, \PW\ ranges from
0 to 1. Typically, one measures the full-width-at-half-maximum
(FWHM). However, the FWHM is not a robust measure of the pulse signal width for 
our application, which may deals with the RFI and signals with low \SNR. In 
order to robustly
measure the width, we first fit the pulse profile to multiple Gaussian
components \citep{KWJ94}. Overlapping components are combined. \PW\
is then calculated using an amplitude-weighted sum of the FWHM of each
component.  The width is thus a useful parameter to
discriminate pulsar candidates from RFI, since measured pulsar pulse widths are 
usually less than 10\%
\citep{Rankin83, LM88, MGR11} and RFIs usually result in broad waveforms.  
Admittedly this breadth can be comparable to that
seen in millisecond pulsars, however.  

\item The persistence of the signal in the time domain (\IT).

The persistence of the signal in the time domain is a measure of the fraction of 
the observation in which the candidate signal is present. The candidate file 
usually contains a three dimensional data cube, i.e. the signal intensity as the 
function of the time index, the frequency index, and the pulse phase index.  
From the candidate file, \PP\ reads in the folded pulse profile for each 
sub-integration (i.e. the pulse profile at each time index) and then
calculates the on- and off-pulse amplitude ratio $r_{\rm T}$: \begin{equation}
	r_{ {\rm T}}=\frac{\textrm{Average of signal level in the pulse window}} { 
	\textrm{Average of signal level outside the pulse window} }\,,
\end{equation}
where the on-pulse window is defined as that region of pulse phase that lies 
within the FWHM region of each profile component and the off-pulse window covers 
the remainder. Using $r_{\rm T}$ calculated for each sub-integration, \PP\ then 
computes \IT:
\begin{equation}
	\textrm {\IT}=\frac{\textrm{Number of sub-integrations with } r_{ {\rm T}
	}>\alpha_{\rm T} }{ \textrm{Total number of sub-integrations}}\,.
	\label{eq:pertt}
\end{equation}
Here $\alpha_{\rm T}$ is a preset threshold, whose default value is $1$ in \PP.  
By definition, $\eta_{\rm T} \in [0,1]$. Since the pulsar signal is expected to 
persist for most of the observing session, true pulsar signals should have a 
high value of $\eta_{\rm T}$.

\item The persistence of the signal in the radio frequency domain (\IF).

The persistence of the signal in the radio frequency domain is a measure of that 
fraction of the bandwidth in which the candidate signal is present. Similar to 
the calculation of the $r_{\rm T}$, \PP\ reads in the folded pulse profile for 
each sub-band, then computes the on-and-off pulse amplitude ratio $r_{\rm F}$.  
\IF\ is then computed as \begin{equation}
	\textrm {\IF}=\frac{\textrm{Number of sub-bands with } r_{ {\rm 
	F}}>\alpha_{\rm F}
	}{ \textrm{Total number of sub-bands}}\,,
	\label{eq:pertft}
\end{equation}
where the threshold $\alpha_{\rm F}$ is set to a default value of $1$ in \PP. As 
with $\eta_{\rm T}$, $\eta_{\rm F}\in[0,1]$. Since the pulsar signal is expected 
to be broadband, true pulsar signals should have a high value of $\eta_{\rm F}$.

\item The ratio between the pulse width and the DM smearing time (\ID).

\PP\ reads the barycentric 
period ($p_{\rm bar}$), the frequency channel width ($\Delta f_{\rm c}$), the 
center frequency ($f$), and the dispersion measure ($DM$) from the candidate 
file, then calculates the dispersive smearing time across a single frequency 
channel:
\begin{equation}
	\Delta\tau=8.3 \, {\rm \mu s} \left(\frac{\Delta f_{\rm c}}{\rm MHz}\right) 
	\left(\frac{f}{\rm GHz}\right)^{-3} \left(\frac{DM}{\rm cm^{-3} pc}\right)\,.
\label{eq:dmsm}
\end{equation}
Together with the fractional pulse width $w$, \ID\ is calculated as 
\begin{equation}
	\eta_{\rm DM}=\frac{p_{\rm bar} w}{\Delta \tau}\,.
	\label{eq:dmint}
\end{equation}
Since the pulse width of any true astronomical signal must be 
greater than the dispersive smearing time across a single frequency channel, 
\ID\ is expected to be greater than 1 for a real pulsar signal. Due to the small 
difference between the $p_{\rm bar}$ and the $p_{\rm top}$, either can be used 
in the above calculation. 

This simple definition of DM score neglects the effects of sampling time. This 
does not have a significant impact on the \PP\ ranking, because the sampling 
time effect only increases the measured width \PW\ of a pulse profile, which 
increases the ranking score rather than decreasing it.

\end{enumerate}

\subsection{Scores}
\label{sec:like}
\begin{figure} \centering \includegraphics[totalheight=2.7in]{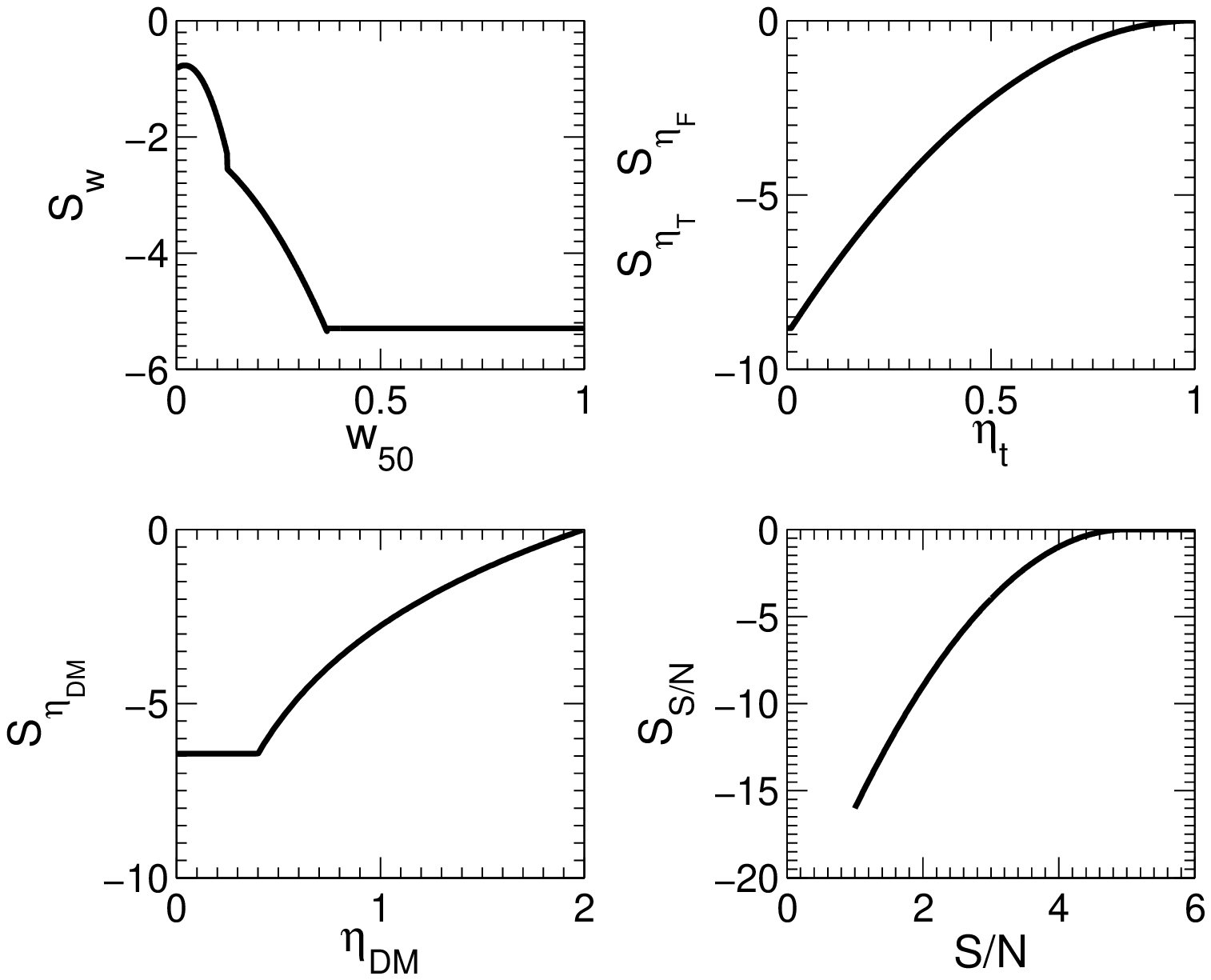} \caption{ The 
	functions $S_{\textrm{\SNR}}, S_{\textrm{\PW}}, S_{\textrm{\IT}}, 
	S_{\textrm{\IF}}$,and $S_{\textrm{\ID}}$ used by \PP\ to determine the overall 
	candidate score. }
	\label{fig:lik} \end{figure}

\begin{figure} \centering \includegraphics[totalheight=2.3in]{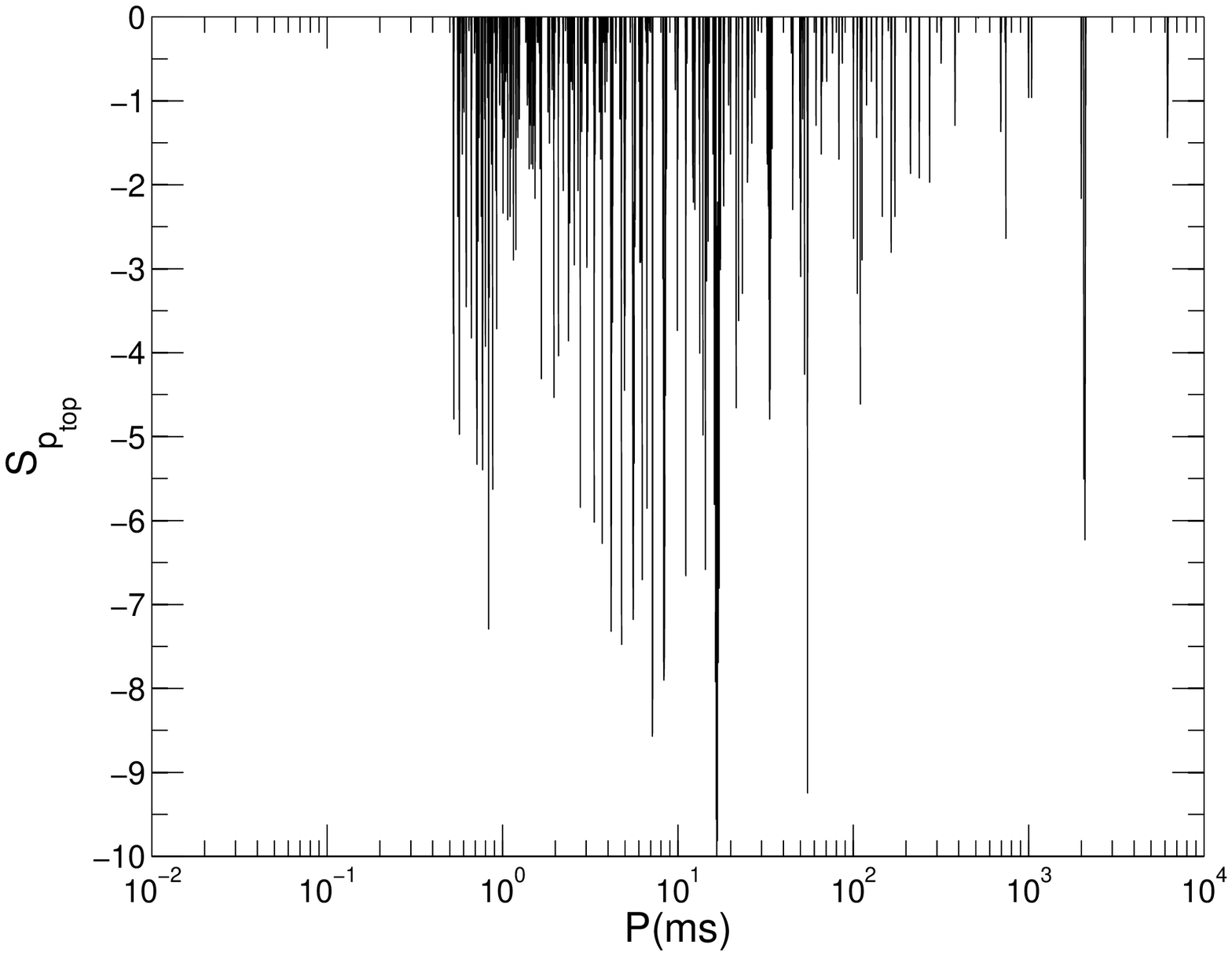} \caption{ 
		$S_{\textrm{\PT}}$ for the \GBN\ survey. The x-axis is the topocentric 
		period in milliseconds, and the y-axis is the value of $S_{\textrm{\PT}}$. }
	\label{fig:qp} \end{figure}

\PP\ combines the measured quality factors and calculates a final score, $S$,
which is stored in the database and later used by the ARCC Explorer to sort the 
candidates.  The final score is defined as a linear combination of individual 
ones, i.e.
\begin{eqnarray}
	S&=& \beta_{\textrm{\SNR}} S_{\textrm{\SNR}}(\textrm{\SNR})+ 
	\beta_{\textrm{\PT}} S_{\textrm{\PT}}(\textrm{\PT}) \nonumber \\
	&&+\beta_{\textrm{\PW}} S_{\textrm{\PW}}(\textrm{\PW}) + \beta_{\textrm{\IT}} 
	S_{\textrm{\IT}}(\textrm{\IT})\nonumber\\
	&&+ \beta_{\textrm{\IF}} S_{\textrm{\IF}}(\textrm{\IF}) + \beta_{\textrm{\ID}} 
	S_{\textrm{\ID}}(\textrm{\ID})\,,\label{eq:totlike}
\end{eqnarray}
where $\beta_{\textrm{\SNR}}, \beta_{\textrm{\PT}}, \beta_{\textrm{\PW}}, 
\beta_{\textrm{\IT}}, \beta_{\textrm{\IF}}$, and $\beta_{\textrm{\ID}}$ are 
constants with default values of $1$. The functions $S_{\textrm{\SNR}}, 
S_{\textrm{\PW}}, S_{\textrm{\IT}}, S_{\textrm{\IF}}$,and $S_{\textrm{\ID}}$ are
\begin{eqnarray}
	S_{\textrm {\SNR}}(\textrm{\SNR}) &=& \left\{ \begin{array}{ll}
		-(\textrm{\SNR}-5)^2\, & \text{for \SNR $\le 5$,} \\
		0\, & \text{for \SNR $> 5$.} \\
	\end{array} \right.
	\,\\
	S_{\textrm {\PW}}(\textrm{\PW}) &=& \left\{ \begin{array}{c}
		-280.7 \textrm{\PW}^2+11.4 \textrm{\PW}-1.6\,,		 \\
		 \text{for \PW$<0.125$\,,} \\
		-37.9 \textrm{\PW}^2-4.1 \textrm{\PW}-4.0 \,,\\
		\text{for $0.125\le$\PW $< 0.6$} \,,\\
		-20 \text{ for $0.6\le$\PW} \,,\\
	\end{array} \right. \,\\
	S_{\textrm {\IT}}(\textrm{\IT}) &=& -9\left(\textrm{\IT}-1\right)^2\,,\\
	S_{\textrm {\IF}}(\textrm{\IF}) &=& -9\left(\textrm{\IF}-1\right)^2\,,\\
	S_{\textrm {\ID}}(\textrm{\ID}) &=& \left\{ \begin{array}{ll}
							-10.2 &\text{ \ID$<0.4$,} \\
							-4(\textrm{\ID}-2)^2 &\text{$0.4 \le$ \ID $\le$2,} \\
							0 &\text{$2<$\ID. }
	\end{array} \right.\,
\end{eqnarray}
Each of these functions is shown in \FIG{fig:lik}. The form of these functions 
was inspired by the natural logarithm of the probability distribution of the 
relevant quality factor. For example, the function $S_{\textrm{\PW}}$ is an 
analytic approximation to the natural logarithm of the pulse width probability 
distribution determined from all radio pulsars in the ATNF catalog 
\citep{MHT05}. The above functions are not tuned to a particular survey.
However, the function $S_{\textrm{\PT}}$, which characterizes the closeness of 
the candidate period to any RFI signal's value, has to be determined according 
to the local RFI environment of the survey under study.  At the beginning of a 
survey, one does not have any information about the RFI.  In that case, the 
$S_{\textrm{\PT}}$ is chosen to be $0$. As the number of candidates becomes 
large, e.g. $10^4$, one can start to construct $S_{\textrm{\PT}}$ using the 
following recipe: i) Calculate
the histogram of the periods of all the candidates; ii) Use a running-median 
filter to determine the baseline in the histogram; iii) Remove the baseline from 
the histogram; iv) Take the negative logarithm of the histogram and re-scale it 
so that it ranges from -10 to 0; v) Interpolate the resulting distribution in 
order to create the analytic continuous function $S_{\textrm{\PT}}$. Using data 
to generate scores is similar to the ideas of machine-learning based techniques, 
where we use the data themselves as the training set to generate the RFI scores.  
In the released \PP\ package, there is a dedicated tool, \emph{buildSP}, which 
can generate this function from a list of candidate periods. As an example, 
$S_{\textrm{\PT}}$ for the \GBN\ survey is shown in \FIG{fig:qp}.

\begin{figure} \centering \includegraphics[totalheight=2.3in]{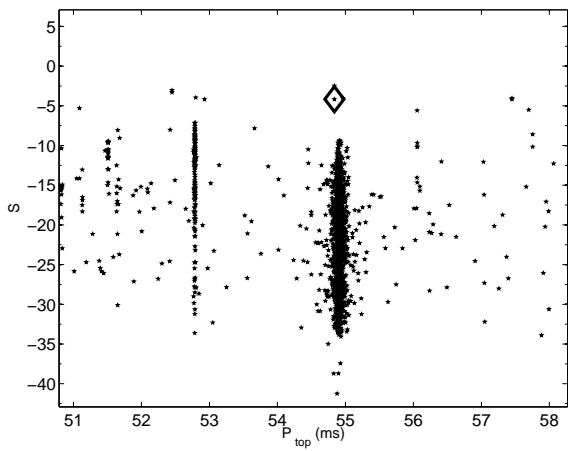} \caption{ 
		The \PP\ score and periods for a subset of candidates. The dots are all 
		identified as RFIs, while the diamond symbol is the candidate later 
		identified to be a pulsar. Although the periods are similar, the \PP\ score 
		of the candidate is still higher than the scores of RFI. }
	\label{fig:rfi} \end{figure}

The use of $S_{\textrm{\PT}}$ was inspired by the `birdie-list zapping 
technique' (e.g. \citealt{LK05}), where one removes all of the candidates with 
$p_{\rm top}$ in the period range where RFI often appears\footnote{There is a major difference between the birdie-list and the period scoring technique. The birdie-list is usually used before the harmonic summing, while the $S_{\textrm{\PT}}$ is applied to the final candidates.}. Instead of completely 
removing such candidates, \PP\ only reduces their final score. If other 
qualities are good, \PP\ can still rank such a candidate highly. This allows for 
the discovery of pulsars with periods similar to the local RFI. A particular 
example is shown in \FIG{fig:rfi}: although the period of the candidate is close 
to the RFI, \PP\ still gives the candidate a high final score to make the 
candidate stand out against the RFI because of the other scores.

\section{Implementation and Application}
\label{sec:imp}

\subsection{Implementation}

The source code for \PP\ is located at
http://sourceforge.net/projects/pulsareace/. Currently, there
exist three versions of \PP\ where the only difference between these is
the type of candidate file it analyzes. One version analyzes PRESTO
candidate files (.pfd), one for pdmp \citep{HSM04} candidate files
(.ar), and one for image plots (.png). For
.png files, \PP\ scans user-defined regions of the image to
determine the pulse profile, sub-integrations, and sub-band data. These
regions can be specified in the command line with the pixel coordinates
of the left, right, top, and bottom of each panel. The idea behind the .png
version of \PP\ is to extend its ability to analyze candidates generated
in alternative pipelines, where the users measure the geometry parameters just 
once and they can process all the image with an identical command
line. The values of the candidate period and DM are entered via command-line 
arguments upon execution. If these are not given, \PP\ ignores
$S_{\textrm{\PT}}$ and $S_{\textrm{\ID}}$, when calculating the score.

As discussed before, \PP\ uses several statistical thresholds together with 
several preset constants (i.e. $\alpha_{\rm T}, \alpha_{\rm F}, 
\beta_{\textrm{\SNR}}, \beta_{\textrm{\PT}}, \beta_{\textrm{\PW}}, 
\beta_{\textrm{\IT}}, \beta_{\textrm{\IF}}$, and $\beta_{\textrm{\ID}}$) to 
calculate the score. \PP\ allows the user to override each of these parameters 
via command-line arguments. For example, one can increase $\beta_{p_{\rm top}}$ 
for data with comparatively worse RFI.

\subsection{Evaluation}
ARCC students have
visually inspected over $10^5$ candidates from a part of the \GBN\ survey and 
have identified over 70 confirmed pulsars (including previous
discoveries). This data set is complete and un-biased, because all the
candidates have been visually inspected at least once. Such a unique data set is 
very valuable to evaluate the effectiveness of
automatic pulsar ranking systems such as \PP. 

\begin{figure*} \centering \includegraphics[totalheight=4.5in]{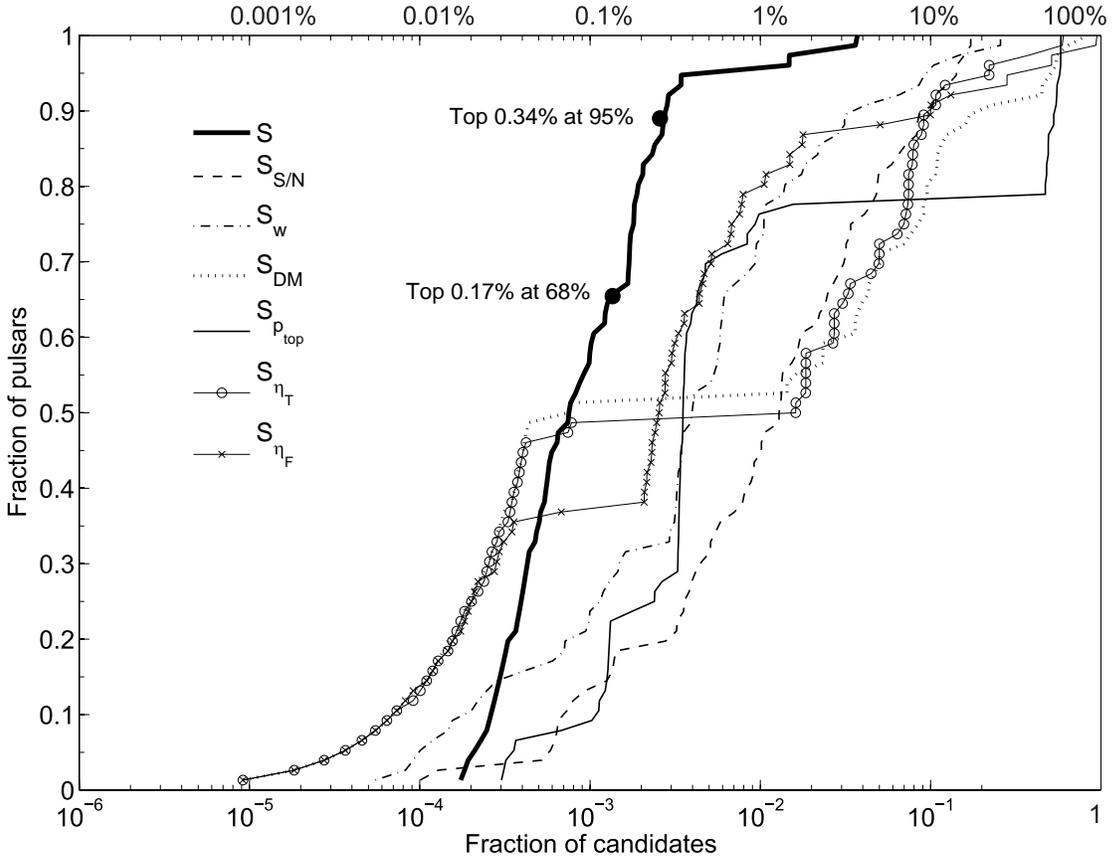}
\caption{In order to determine the efficiency of \PP, candidates are
sorted in descending order according to the score generated by \PP. In
the above figure, the fraction of top ranked candidates is plotted versus the 
fraction of confirmed pulsars in
those candidates. The thick solid, dashed, dot-dashed, dotted, solid, circled, 
and cross-marked curves are for the ranking using the score $S, 
S_{\textrm{\SNR}}, S_{\textrm{\PW}}, S_{\textrm{DM}}$, $S_{\textrm{\PT}}$, 
$S_{\textrm{\IT}}$, and $S_{\textrm{\IF}}$, respectively. We can see that using 
the final score (S) is better than using the individual ones. The ranking using 
the final score $S$ puts 68\% of confirmed pulsars
in the top \FIRSTL\ of ranked candidates, 95\% are in the top \SECL,
and all are in the top \THRL. } \label{fig:roc} \end{figure*}

The ideal candidate sorting algorithm would score all real pulsars
higher than all other candidates. Therefore, the distribution of all the known
pulsars in the list of candidates ranked by score is a good measure
of the effectiveness of the ranking algorithm. Here, we used the pulsars, which 
are identified by ARCC students in \GBN\ survey, as tracers to evaluate the \PP\ 
performance.
The measured detection rate of \PP\ is shown in \FIG{fig:roc}, where 68\% of 
confirmed pulsars are in the
top \FIRSTL\ of \PP-ranked candidates, 95\% are in the top \SECL,
and all are in the top \THRL. These results indicate that 
inspecting
candidates in order of decreasing \PP\ score will significantly increase the
rate of pulsar identification. For this evaluation, the statistical thresholds 
and preset constants were set to their default values as described above.

For most of the time, the processing speed of \PP\ is limited by the time of 
reading in files. An Intel\textregistered\ 2.4 GHz processor with 6 Mb cache 
usually processes a candidate file within 100 to 500 ms, which corresponds to 
0.2 to 1 million candidates per processor per day. 

\section{Discussions and Conclusions}
\label{sec:con}

In this paper, we have described \PP, a software package for post-analysis 
processing of pulsar survey candidates. \PP\ uses a set of algorithms to analyze 
a pulsar candidate and calculate a score, which is a measure of how likely a 
candidate is to be a real pulsar. These algorithms are described and the 
effectiveness of \PP\ has been evaluated. Using candidates generated by the \PP\ 
survey and
inspected by students in the ARCC program, it was shown that \PP\ significantly 
increases the rate of identifying pulsars. For example, four million candidates 
requires approximately $10^3$ person hours in order to visually inspect each 
one.  If one pre-sorts these candidates according to their \PP\ score, 100\% of 
the pulsars are expected to be in the top 150 thousand candidates. Inspecting 
these candidates only requires 40 person hours. Such efficiency will, hopefully, 
help the
pulsar surveys using future large telescopes, such as the Five-hundred-meter 
Aperture Spherical Radio Telescope (\citealt{NWZZJG04, SLKMSJN09}) and the 
Square Kilometre Array (\citealt{KS10, SKSL09}).

\PP\ uses six quality factors to determine a candidate's score. These are
the signal-to-noise ratio (\SNR), the candidate period, the pulse width,
the signal's persistence in the time and the frequency domains, and the pulse
width to DM smearing time ratio. These quality factors are chosen because
they are readily available from standard pulsar searching pipelines
(e.g. PRESTO and SIGPROC). Also, human experience has shown that these
particular quality scores are helpful in differentiating between pulsar
candidates and RFI. As shown in \FIG{fig:roc}, there is no single quality factor 
dominating the final score. For example, typical survey analyses will remove 
candidates whose periods lie within RFI-dominated regions (e.g.  the 
`birdie-list zapping technique'). \PP\ simply reduces the score of such 
candidates. Thus, it is still possible to find pulsars within the RFI-dominated 
regions. 

Beside the six quality factors used in this paper, there are other 
possibilities. For example, \citet{EAT09, EMK11} find that the $\chi^2$ of fit 
to the theoretical DM-\SNR\ curve and other factors can be useful for neural 
network algorithms in identifying pulsars. From our experience with \PAF\ and 
HTRU North surveys, the score function for the DM-\SNR\ curve fitting $\chi^2$ 
will not be a simple shape. In this way, a more complex scoring scheme with more 
quality factors may further improve the current performance of \PP. 

\PP\ is fully pre-determined and does not require any training data sets.  This 
is different from other approaches
that use neural networks \citep{EAT09, EMK11, BB12}, which determine the 
strategy for ranking candidates by `learning' knowledge from training data sets.  
Although \PP\ does not require such initial training data sets, it can be 
further fine-tuned when such data become available.  

As a caveat, using \PP\ introduces selection effects in the searching.  For
machine-learning algorithms, it is hard to quantify the selection effects, since 
they are inherited from the training data sets. For \PP\, we know exactly what 
the selection effects are and the users can adjust the score weights to adapt to 
particular purposes. As indicated in \FIG{fig:lik}, \PP\ prefers candidates with 
small pulse widths, high \SNR, wide-band signals, and persistent pulses. It also 
down-weights the candidates with pulse profiles narrower than the channel DM 
smearing widths.  Using multiple scores reduces the chance of missing good 
candidates to a certain degree, although it may still give low scores for the 
candidates with wide pulse profiles and low \SNR. For certain pulsars, the pulse 
energy can decrease by a factor of ten or more over a short timescale and then 
increase just as sharply afterwards \citep{Bac70}. Such `nulling' pulsar may get 
a lower score due to the persistence score \IT. Similarly, in the radio 
frequency domain, signals can scintillate due to the interstellar medium 
\citep{Rick90}, which reduces the score \IF.  

\PP\ can be a starting point for machine-learning algorithms. Since we have 
demonstrated that the quality factors calculated by \PP\ effectively quantify 
how likely a candidate is of being a pulsar, more advanced algorithms may 
further improve \PP's scoring performance.

\section{Acknowledgement}

K.~J.~Lee gratefully acknowledges support from ERC Advanced Grant
``LEAP'', Grant Agreement Number 227947 (PI Michael Kramer). K.~Stovall,
F.~Jenet,  and the students in the ARCC program acknowledge the support
from NSF AST 0545837 \& 0750913. Pulsar research at UBC is supported by an
NSERC Discovery Grant and Discovery Accelerator Supplement, by CANARIE and
by the Canada Foundation for innovation. The work of CB, MR, JF, AW, SB,
and XS was partially supported by the Office for Undergraduate Students at
the University of Wisconsin -- Milwaukee and the NSF through CAREER award
number 0955929, PIRE award number 0968126, and award number 0970074. PL 
acknowledges the support of IMPRS Bonn/Cologne and NSERC PGS-D
We thank R.~Eatough, D.~Nice, and J.~P.~W.~Verbiest for reading the
manuscript and for the helpful comments. % and illuminating discussions.

\label{lastpage}

%\bibliographystyle{mn2e}
%\bibliography{ms}
%\clearpage

\end{document}